\documentstyle[12pt]{article}

\newcommand{\be}{\begin{equation}}
\newcommand{\ee}{\end{equation}}
\newcommand{\bea}{\begin{eqnarray}}
\newcommand{\eea}{\end{eqnarray}}

\newcommand{\RR}{\rangle}

\newcommand{\cd}{\cdots}

\makeatletter
\newdimen\normalarrayskip              
\newdimen\minarrayskip                 
\normalarrayskip\baselineskip
\minarrayskip\jot
\newif\ifold             \oldtrue            

\makeatother
\newlength{\extraspace}
\newlength{\extraspaces}
\input{tcilatex}

\begin{document}

\addtolength{\baselineskip}{3mm}

\thispagestyle{empty}

\begin{flushright}
\baselineskip=12pt
quant-ph/9905024\\
\hfill{  }\\ May 1999
\end{flushright}
\vspace{.5cm}

\begin{center}
\baselineskip=24pt

{\Large {\bf {No Signalling and \\[0pt]
Probabilistic Quantum Cloning}}}\\[15mm]

\baselineskip=12pt

{\bf Lucien Hardy\footnote{{\bf E-mail: {\tt l.hardy1@physics.ox.ac.uk}}}
and David D. Song\footnote{{\bf E-mail: {\tt d.song1@physics.ox.ac.uk}}}} \\[%
8mm]
{\it Department of Physics, Clarendon Laboratory\\[0pt]
Parks Road, University of Oxford\\[0pt]
Oxford OX1 3PU, U.K.} \vspace{5cm}

{\sc Abstract}

\begin{minipage}{15cm}
\baselineskip-12pt
We show that the condition of no faster-than-light signalling restricts
the number of quantum states that can be cloned in a given Hilbert space.
This condition leads to the constraints on a probabilistic quantum
cloning machine (PQCM) recently found by Duan and Guo.

\end{minipage}
\end{center}

\vfill

\newpage

\pagestyle{plain} \setcounter{page}{1}

J.S. Bell showed, in his famous paper \cite{bell}, the non-local nature of
quantum mechanics. Nevertheless, this non-locality cannot be used to
communicate faster-than-light due to indistinguishability between two
mixtures. On the other hand, if quantum states can be cloned perfectly,
special relativity would inevitably be violated. After Wootters and Zurek
had shown \cite{zurek} that nonorthogonal quantum states cannot be cloned,
various authors \cite{buzek, massar, bruss} showed that the approximate
cloning is possible, i.e. with fidelity less than one. Gisin demonstrated
\cite{gisin} that the imperfect cloning still does not violate special
relativity by showing the bounds on the fidelity of QCM for the no-signalling
constraint and the fidelity of the approximate QCM are equal. Recently, Duan
and Guo  (DG) showed \cite{duan1,duan2} that linearly independent quantum states
can be cloned perfectly sometimes, a probabilistic quantum cloning. In this
paper, we show that the conditions on no faster-than-light signalling lead to
the constraints of DG on probabilistic quantum cloning machines (PQCM).

Consider arbitrary, but not necessarily orthogonal, states for Bob,
$|B_{n}\rangle $ where $n=1$ to $N$, in a Hilbert space of dimension
$N$.  Bob's states can be realised by Alice's measurements
 on the following entangled states.
 \begin{equation}\label{one}
|\psi \rangle _{AB}=\frac{1}{\sqrt{N}}\sum_{n=1}^{N}  |n\rangle _{A}|B_{n}\rangle 
\end{equation}
where $|n\RR_A$ are orthogonal states.
We will consider two possible measurements by Alice.  She can either
measure onto the $|n\rangle_A$ basis (call this measurement $A_1$) or
onto some other basis (call this measurement $A_2$).  When she measures
$A_1$ the states  $|B_{1}\rangle ,\cdots
,|B_{N}\rangle $ are prepared for Bob and when she measures $A_2$ she
prepares a different set of states which we will call $|B_{N+1}\rangle ,\cdots
,|B_{2N}\rangle $ for Bob.  It is clear from (\ref{one}) that the
vectors $|B_{N+1}\rangle ,\cdots ,|B_{2N}\rangle $ are linearly
dependent on the vectors $|B_{1}\rangle ,\cdots ,|B_{N}\rangle $.
Furthermore, it is also clear that one of the second set of states,
$|B_{N+1}\rangle$ say, can be any linearly dependent state.
Let us suppose Bob has a PQCM such that
\begin{equation}
{\rm Input}\;|B_{\alpha}\rangle \stackrel{{\rm prob}\neq 0}{\longrightarrow }%
{\rm Output}\;|B_{\alpha}\rangle ^{\otimes \mu}
\end{equation}
where $\mu$ is large and we assume this PQCM works for some subset of $\alpha
=1,\cdots ,2N$.  Furthermore, the PQCM is such that when it succeeds in
producing $\mu$ copies Bob knows that it has succeeded.  He can disregard
those cases where it does not succeed.  Note that if Bob has a PQCM which only makes a single
clone with some non-zero probability then repeated application of this
machine will lead to one which makes many copies though with a much
smaller probability.

Assume that Alice and Bob share a large number of pairs of entangled
pairs $|\psi\rangle_{AB}$.
Let us imagine that Alice wants to communicate a 0 or a 1 to Bob.  If
she wants to communicate a 0 (1) she measures $A_1$ ($A_2$) on all her
particles.  Bob can now use his PQCM to attempt to infer which
measurement Alice made.  Since sufficiently many
entangled pairs are shared in each group, Bob's PQCM is guaranteed to
produce $\mu$ (which is a large number) copies for at least one of the
pairs.  Consider one such pair where Bob has been successful in producing
$\mu$ copies of particle $B$.  Bob can attempt to establish what the
state of these copies is by making appropriate measurements.
Suppose Bob's PQCM can in fact only clone
$N+1$ states of $|B_n\rangle$, say $|B_1\RR,\cd,|B_{N+1}\RR$.
For inputs $|B_{k}\rangle$ where $k=N+2$ to $2N$ (which the PQCM
cannot clone) the general output state can be written
\be\label{general}
\sum_{l=1}^{N+1} c_l^k |B_l\RR^{\otimes\mu} + d^k |\phi^k\RR
\ee
where $l=1$ to $N+1$ and where the state $|\phi^k\rangle$ has zero overlap with
$|B_l\RR^{\otimes\mu}$ for $l=1$ to $N+1$.  To attempt to establish the
state of the clones Bob divides them up into $N+1$ groups and
makes projective measurements on each clone onto the state
$|B_l\rangle$ for the $l$th group.  If all the clones in the $l$th group
successfully project onto $|B_l\RR$ then Bob can conclude that
output state of the $\mu$ clones behaved as state
$|B_l\rangle^{\otimes\mu}$.   In such a case he will place a $\surd$ in
the $|B_l\rangle^{\otimes\mu}$ column.  If none of the groups have this
property then he can place a $\surd$ in the $|\phi\RR$ column.  It is
very unlikely that more than one group will have this property (but if
this does happen we can put a $\surd$ in the $|\phi\rangle$ column
instead).  The
idea that the state may behave as $|B_l\rangle^{\otimes\mu}$ is
motivated by the fact that we can write any general output state down as
in (\ref{general}).  However, we should be a little careful with this
equation.  It is possible that terms other than
$|B_l\rangle^{\otimes\mu}$ could lead to a $\surd$ in column $l$.
Equation (\ref{general}) should only be regarded as motivating the idea
that the state may behave as one of $|B_l\rangle^{\otimes\mu}$.

Let us consider the most general case as shown in the  table 1.
\begin{center}
{\rm Output}

{\rm Input} 
\begin{tabular}{|l||l|l|l|l||l|l|}
\hline
& {$|B_1\rangle^{\otimes \mu}$} & {$\cdots$} & {$|B_{N-1}\rangle^{\otimes \mu}$}
& {$|B_N\rangle^{\otimes \mu}$} & {$|B_{N+1}\rangle^{\otimes \mu}$} &
{$|\phi\RR$} \\ \hline\hline
{$|B_1\rangle$} & {$\surd$} &  &  &    &  & \\ \hline
{$\vdots$} &  & {$\ddots$} &  &  &  &   \\ \hline
{$|B_{N-1}\rangle$} &  &  & {$\surd$} &  &  &    \\ \hline
{$|B_N\rangle$} &  &  &  & {$\surd$} &  &   \\ \hline\hline
{$|B_{N+1}\rangle$} &  &  &  &  & {$\surd$} &    \\ \hline
{$|B_{N+2}\rangle$} & {$\surd$} &{$\cdots$}  &{$\surd$}  &{$\surd$}  &{$\surd$}   &{$\surd$}  \\ \hline
{$\vdots$} &{$\vdots$}  & {$\ddots$} &{$\vdots$}  &{$\vdots$}  &{$\vdots$}  &{$\vdots$} \\ \hline
{$|B_{2N}\rangle$} &{$\surd$}  &{$\cdots$}  & {$\surd$} &{$\surd$} &{$\surd$}   &{$\surd$}  \\ \hline
\end{tabular}

{\rm Table  1}
\end{center}

This table shows all the possible places $\surd$'s could be placed.
For the inputs which can be cloned there can only be one $\surd$ in the
corresponding row.  For the other cases there could be a non-zero
probability for a $\surd$ in any of the columns including the $|\phi\RR$
column.

Now let us consider how Alice may try to communicate faster-than-light
to Bob.  Take a case in which Bob's cloning machine tells him that he
has successfully cloned $\mu$ copies of particle $B$ (of course this will
only actually be true if the input was one of $|B_1\RR$ to $|B_{N+1}$).
Define $P(n|A_i)$ as the probability that Bob places a $\surd$ in the
$|B_l\rangle^{\otimes\mu}$ column given that Alice measured $A_i$.
Bob can proceed in the following way.  If he places a $\surd$ in a column
corresponding to $l=1$ to $N$ he guesses that Alice sent a 0 and if he
places a $\surd$ in column $N+1$ he guesses that Alice sent a 1.  The
conditional probabilities for his inferring a 0 or a 1 can be read off
from the table.  They are
\be
P_0(A_1)\equiv\sum_{n=1}^N P(n|A_1)=1, \; \; P_1(A_1)\equiv P(N+1|A_1) =0
\ee
\be
P_0(A_2)\equiv\sum_{n=1}^N P(n|A_2) \leq 1-P(N+1|A_2),
\; \; P_1(A_2)\equiv P(N+1|A_2)\not= 0
\ee
Although Bob will sometimes guess wrongly, if this process is repeated
many times Bob can be sure of guessing correctly.

We know that faster-than-light signalling is not possible and hence it
cannot be possible to clone more than $N$ states.  Note further, that we
can also see from this argument that the states which can be cloned must
all be linearly independent.  To see this imagine that the first $N-1$
of the states $|B_n\rangle$ are linearly independent and the $N$th is
linearly dependent on these.  Then we could run the whole argument again
replacing $N$ by $N'=N-1$ throughout and again arrive at a
contradiction.
Duan and Guo's  PQCM \cite{duan1,duan2} has the constraint that only linearly independent
quantum states can be
cloned perfectly with non-zero probability. Therefore the condition for no signalling yields the
constraint on a PQCM.

This proof that the no faster-than-light signalling constraint leads to
the constraint on probabilistic cloning is stronger in some respects
than Gisin's proof relating to deterministic cloning since we do not
make any assumptions about the nature of the output state of the cloning
machine (other than that it successfully clones in those cases we
require it to successfully clone).

{\bf Acknowledgements}.  We would like to thank Dagmar Bruss for
correspondence on this topic.


\begin{thebibliography}{9}
\bibitem{bell}  J.S. Bell, Physics 1 (1964) 195.

\bibitem{zurek}  W.K. Wootters and W.H. Zurek, Nature 299 (1982) 802

\bibitem{buzek}  V. Buzek and M. Hillery, Phys. Rev. A 54 (1996) 1844

\bibitem{massar}  N. Gisin and S. Massar, Phys. Rev. Lett. 79 (1997) 2153

\bibitem{bruss}  D. Bruss, D. P. DiVincenzo, A. Ekert, C. Macchiavello and
J.A. Smolin, Phys. Rev. A57 (1998) 2368

\bibitem{gisin}  N. Gisin, Phys. Lett. A242 (1998) 1

\bibitem{duan1}  L.-M. Duan and G.-C. Guo, Phys. Rev. Lett. 80 (1998) 4999

\bibitem{duan2}  L.-M. Duan and G.-C. Guo, quant-ph/9705018

\end{thebibliography}
\end{document}